\documentclass[useAMS,usenatbib]{mn2e}
\usepackage{graphicx}
 
\def\be{\begin{equation}} 
\def\ee{\end{equation}}

\def\HI{\hbox{H~$\scriptstyle\rm I\ $}} 
\def\HII{\hbox{H~$\scriptstyle\rm II\ $}}

\def\gsim{\lower.5ex\hbox{\gtsima}} 
\def\lsim{\lower.5ex\hbox{\ltsima}} \def\gtsima{$\; \buildrel > \over 
\sim \;$} \def\ltsima{$\; \buildrel < \over \sim \;$} \def\prosima{$\; 
\buildrel \propto \over \sim \;$} \def\gsim{\lower.5ex\hbox{\gtsima}} 
\def\lsim{\lower.5ex\hbox{\ltsima}} 
\def\simgt{\lower.5ex\hbox{\gtsima}} 
\def\simlt{\lower.5ex\hbox{\ltsima}} 
\def\simpr{\lower.5ex\hbox{\prosima}} \def\la{\lsim} \def\ga{\gsim}

\def\gtsima{$\; \buildrel > \over \sim \;$} 
\def\ltsima{$\; \buildrel < \over \sim \;$} 
\def\gsim{\lower.5ex\hbox{\gtsima}} 
\def\lsim{\lower.5ex\hbox{\ltsima}} 
\def\simgt{\lower.5ex\hbox{\gtsima}} 
\def\simlt{\lower.5ex\hbox{\ltsima}} 
\def\simpr{\lower.5ex\hbox{\prosima}} 
\def\la{\lsim} 
\def\ga{\gsim}

\def\E3{{\cal E}_{\rm g}^{III}}

\title[LAE-LBG connection]{Ly$\alpha$ Emitters and Ly-Break Galaxies: dichotomous twins?} 
\author[Dayal \& Ferrara]{Pratika Dayal$^{1}$\thanks{E-mail: 
dayal@aip.de (PD)}\& Andrea Ferrara$^{2}$    \\ 
$^{{1}}$ Leibniz Institute for Astrophysics , An der Sternwarte 16, 14482 Potsdam, Germany\\ 
$^{2}$ Scuola Normale Superiore, Piazza dei Cavalieri 7, 56126 Pisa, Italy}

\begin{document} 
 
\date{} 
 
\pagerange{\pageref{firstpage}--\pageref{lastpage}} \pubyear{2002} 
 
\maketitle 
 
\label{firstpage} 
\begin{abstract}
We extend our previous studies aimed at modeling Lyman Alpha Emitters (LAEs) to the second population of high redshift sources, Lyman Break Galaxies (LBGs), with the final goal of investigating the physical relationship between them. We use a set of large ($\approx 10^6 {\rm Mpc}^3$) cosmological SPH simulations that include a detailed treatment of star formation, feedback, metal enrichment and supernova dust production; these same simulations have already been shown to successfully reproduce a large number of observed properties of LAEs (Dayal et al. 2010). We identify LBGs as galaxies with an absolute ultraviolet (UV) magnitude $M_{UV} \leq -17$, consistent with current observational criteria. We then compute the evolution of their (a) UV Luminosity Function (LF), (b) UV spectral slope, $\beta$, (c) stellar mass function, (d) star formation rate (SFR) density, (e) (specific) Star Formation Rate  (sSFR), and compare them with available data in the redshift range $6 < z < 8$.  With no further parameter tuning, the model reproduces the redshift evolution of the LBG UV LF, stellar mass function, and SFR density extremely well. It predicts a $z$-independent $\langle \beta \rangle \approx -2.2$, in agreement with the most recently updated data sets at $z \approx 6$, but in slight tension with the steeper $\beta-M_{UV}$ observed at $z \approx 7$. The mean LBG sSFR increases from 6.7 Gyr$^{-1}$ at $z \approx 6$ to 13.9 Gyr$^{-1}$ at $z \approx 8$, and it is largest for the smallest  ($M_* < 10^{8.5} {\rm M_\odot}$) LBGs, consistent with the recent findings of Schaerer \& de Barros (2010), and possibly resulting from a downsizing process. From a comparison of the simulated LAE and LBG populations, we find no appreciable differences between them in terms of the mass-weighted stellar masses/ages, SFR, and dust content; only the faintest LBGs with $M_{UV} \geq -18 (-19)$ $z \approx 6 (8)$ do not show an observable Lyman Alpha (Ly$\alpha$) line. LAEs hence represent a luminous LBG subset, whose relative extent depends only on the adopted selection criteria. For example, using the Lyman Alpha equivalent width selection threshold $EW>55$\AA\, of Stark et al. (2010), the LAE fraction increases towards the faintest LBGs. However, for the canonical value of $EW>20$\AA\,, all LBGs with $-23 < M_{UV} < -19$ would be identified as LAEs at $z\approx 6$; the fraction of LBGs showing a Ly$\alpha$ line decreases with increasing redshift from $z \approx 6$ to $8$ due to the combined effects of dust and reionization. We conclude with a brief critical model discussion, which emphasizes the uncertainties inherent to theoretical EW determinations.
\end{abstract} 


\section{Introduction} 

We are in a golden age for the search for high redshift galaxies, located at the beginning of the cosmic dawn. This progress has been made possible by a combination of state of the art instruments and sophisticated selection techniques. The sensitivity of instruments such as the recently installed Wide Field Camera 3 (WFC3) on the Hubble Space telescope (HST), the Subaru and Keck telescopes, have been instrumental in pushing to faint magnitudes, in an effort to search for the first galaxies. This has been complemented by refined selection methods including the dropout-technique introduced by Steidel et al. (1996); the band in which a galaxy `drops-out' of visibility is used as an indicator of the Lyman Break at low redshifts and the break bluewards of the Ly$\alpha$ line at high-redshifts (corresponding to 912 \AA\, and 1216 \AA\, in the rest frame respectively), to estimate the galaxy redshift. Though this method has been extremely useful in identifying high-$z$ LBGs, it has the drawback that the exact source redshift cannot be estimated with complete confidence. An alternate strategy for finding high-$z$ galaxies is the narrow-band technique (e.g. Malhotra et al. 2005; Shimasaku et al. 2006; Kashikawa et al. 2006; Hu et al. 2010) that is based on looking for the relatively unambiguous Ly$\alpha$ emission, and is often combined with a drop-out criterion using broad-band colors (see eg. Ouchi et al. 2008, 2010). The foremost advantage in using such galaxies, called LAEs, is that the observed wavelength of the Ly$\alpha$ line can be used to place stringent constraints on the source redshift. 

We start by summarizing the data collected for LBGs that formed during the first billion years after the Big Bang, i.e. $z \geq 6$. Using NICMOS data from the UDF, ACS and GOODS fields, Bouwens et al. (2007) have identified 627 {\it i}-dropouts, corresponding to $z\approx 6$, down to $M_{UV} \approx -17$. Bouwens et al. (2008) have identified 8 $z \approx 7$ {\it z}-dropout candidates, and have found no $z \approx 9$ $J$-dropouts. Using the recently installed WFC3 on the HST, Oesch et al. (2010) have detected 16 {\it z}-dropouts down to $M_{UV} \approx -18$ between $z =6.5-7.5$. Bouwens et al. (2010a) have used WFC3 data to push the observations to $z=8.0-8.5$ by detecting 5 $Y$-dropouts. Bouwens et al. (2011a) have also found a $z \approx 10$  $J$-dropout; if confirmed, this would be the farthest galaxy known as of date. Using the same WFC3 data but instead, performing a full SED fitting to the optical+infrared photometry of all the high-$z$ galaxy candidates detected at $>5-\sigma$ in at least one of the WFC3/IR broad-band filters, McLure et al. (2010) find a much larger number of candidates, as expected. However, they note that about 75\% of the candidates at $z>6.3$ and 100\% of the candidates at $z>7.5$ allow for a $z<2$ interloper solution. Analyzing the three HUDF and deep wide area WFC3 early release data, Bouwens et al. (2010b) have found 66 candidates at $z=7$ and 47 candidates at $z=8$. Using data from the HAWK-I on the VLT, Castellano et al. (2010) have detected 15 $z \approx 7$ candidates. Finally, again using the HAWK-I, Laporte et al. (2011) have obtained a sample of 10 galaxy candidates at $z \geq 7.5$, lensed by the cluster A2667; of these, they expect at least one, and upto 3 galaxies to be genuine high-$z$ sources. The most solid piece of information that has been collected by such observations is the number density of LBGs as a function of $M_{UV}$, i.e. the LBG UV LF. It is a notable success that the UV LFs collected by these different groups, using various instruments and selection techniques, converge extremely well for any given redshift. 

In spite of this strong observational push, surprisingly little effort has been devoted to theoretically modeling high-$z$ galaxies. Numerical simulations attempting to model galaxy populations and their evolution at $z \geq 6$ are scarce, with the exception of the works by Nagamine et al. (2006), Finlator, Dav\'e \& Oppenheimer (2007), Zheng et al. (2010, 2011) and Forero-Romero et al. (2011). Recently, Salvaterra, Ferrara \& Dayal (2011) have used high resolution simulations implemented with a careful treatment of metal enrichment and supernova (SN) feedback, to get hints on the faint end slope of the LBG LF ($M_{UV} \geq -19.5$) between $z \approx 5-10$ and put constraints on the physical properties of such faint LBGs. The resolution needed to study the faint end, however, meant that only a relatively small cosmological volume of $(10 h^{-1} {\rm comoving \, Mpc; \,cMpc})^3$ could be simulated.

In addition to the nature of these pristine galaxies, an urgent question that needs to be addressed is the relationship between LBGs and LAEs, which remains ambiguous  as a result of the different observational selection techniques used to detect these two high-$z$ galaxy populations. Gawiser et al. (2006) suggest LAEs are less massive and less dusty compared to LBGs, and that LAEs represent the beginning of an evolutionary sequence in which galaxies increase their mass through mergers and dust through star formation. This is in agreement with the work of Pentericci et al. (2007, 2010): although they find LAEs to be less massive and less dust enriched with respect to LBGs, they find no appreciable difference in the age of LBGs with/without Ly$\alpha$ emission. However, this is in contrast with the results found by Kornei et al. (2010); these authors find LAEs to be older, less dusty and lower in star formation compared to LBGs. They conclude that LAEs represent a later stage of galaxy evolution in which SN-driven outflows have have reduced the interstellar medium (ISM) dust covering fraction. Such simplified pictures are complicated by results found by other groups which are now summarized: Finkelstein et al. (2009) find a range of dust extinction, between $E(B-V) = 0.5-4.5$ for 14 LAEs at $z \approx 4.5$. Nilsson et al. (2009) find that LAEs occupy a wide range in color space, implying that not all LAEs are young, pristine objects. Lai et al. (2008) have collected a sample of 70 LAEs at $z\approx 3.1$, about 30\% of which have been detected in the Spitzer $3.6 \mu m$ band. These IRAC (InfraRed Array Camera) detected LAEs are older and more massive than the IRAC undetected sample, which leads these authors to suggest that the IRAC detected LAEs may be a low mass extension of the LBG population. Further, Ouchi et al. (2008) have discussed the overlap of the LAE and LBG UV LFs, pointing out that the overlap between the LFs depends on the the EW criterion used to select these two populations. Along the same lines, Verhamme et al. (2008) propose that at a given redshift, LAEs and LBGs are the same population above a given limiting magnitude; for galaxies fainter than such magnitude, LAEs are less massive compared to LBGs. Finally, Dijkstra \& Wyithe (2011) point out that observationally, LAEs are chosen using sophisticated color-color selection techniques; they caution that simple cuts in EW and UV magnitude can lead to uncertainties in the simulated LAE number densities.

Theoretically too, the picture has remained equally unclear. While considerable efforts have been devoted to using $z \geq 5$ LAEs as probes of reionization and galaxy evolution (e.g. Santos 2004; Dijkstra et al. 2007ab; Kobayashi et al. 2007; Zheng et al. 2010, 2011; Dayal et al. 2008, 2009, 2010, 2011; Dayal \& Ferrara 2011), only scant effort has been spent in establishing a physically motivated connection between LAEs and LBGs at these early epochs. One such effort has been presented in Shimizu \& Umemura (2010) who propose that two kinds of galaxies could be visible as LAEs: early starbursts in young galaxies, or delayed starbursts in evolved galaxies. These authors further claim that LBGs showing a strong Ly$\alpha$ line could possibly include the latter type of LAEs.

In this work, our first aim is to simulate the galaxies that populate the entire observed LBG UV LF at $z \approx 6,7,8$, ranging from galaxies as faint as $M_{UV}\approx -18$ to those as bright as $M_{UV} \approx -23$, to reproduce their observed data sets including the UV LFs, $\beta$, the stellar mass function, the sSFR and the SFR density, thereby pinning down their elusive physical properties. To this aim, we use state of the art cosmological simulations with the same treatment of the metal enrichment and SN feedback, as in Salvaterra, Ferrara \& Dayal (2011). However, to capture the large luminosity range required, we use much larger simulation boxes, such that the volume probed is $(75 h^{-1} {\rm cMpc})^3$. 

Secondly, the simulations used in this work have already been used extensively (Dayal et al. 2009; Dayal, Ferrara \& Saro 2010) to study the nature of $z \geq 6$ LAEs and have successfully reproduced a number of LAE data sets including the Ly$\alpha$ and UV LFs, their spectral energy distributions (SEDs) and Ly$\alpha$ EWs, to name a few. In this work, our aim is to implement this same physically motivated, self-consistent model for all the simulated galaxies at $z \approx 6,7,8$, to identify the galaxies that would be observationally chosen as LAEs and LBGs using the narrow-band and drop-out techniques respectively, at these redshifts. Then, for the first time, we would be able to compare the physical properties of LBGs and LAEs, identified from the same underlying galaxy population, to study the connection between these two galaxy classes in a self-consistent framework.

\begin{table*} 
\begin{center} 
\caption {Average physical properties of LBGs. For each of the redshifts considered in this work (col 1), we show the average stellar mass (col 2), the average SFR (col 3), the average age (col 4), the average stellar metallicity (col 5), the average intrinsic UV spectral slope, i.e. without considering dust attenuation (col 6), the average UV spectral slope including dust attenuation (col 7), the specific SFR (col 8) and the average color excess (col 9).  }
\begin{tabular}{|c|c|c|c|c|c|c|c|c} 
\hline 
$z$ & $\langle M_* \rangle $ & $\langle \dot M_* \rangle$ & $\langle t_* \rangle$ & $\langle Z_* \rangle$ & $\langle \beta_{int} \rangle$ & $\langle \beta \rangle$ & sSFR & $\langle E(B-V) \rangle$\\
$$ & $[{\rm M_\odot}]$ & $[{\rm M_\odot \, yr^{-1}}]$ & $[{\rm Myr}]$ & $[{\rm Z_\odot}]$ & $$  & $$ & $[{\rm Gyr^{-1}}]$ & $$\\ 
\hline
$6$ & $10^{8.6}$ & $3.1$ & $142$ & $0.14$ & $-2.45$ & $-2.2$ & $6.75$ & $0.16$\\
$7$ & $10^{8.3}$ & $2.1$ & $93$ & $0.08$ & $-2.51$ & $-2.28$ & $10.25$ & $0.12$\\
$8$ & $10^{}8.1$ & $2.0$ & $69$ & $0.05$ & $-2.55$ & $-2.32$ & $13.92$ & $0.1$\\
\hline
\label{table1} 
\end{tabular} 
\end{center}
\end{table*} 

\section{Model }
\label{simu}
 
The simulations used in this work have been carried out using a TreePM-SPH code GADGET-2 (Springel 2005) with the implementation of chemodynamics as described in Tornatore et al. (2007); interested readers are referred to Tornatore et al. (2010) for complete details. The adopted cosmological model corresponds to the $\Lambda$CDM Universe with $\Omega_{\rm m }=0.26,\Omega_{\Lambda}=0.74,\ \Omega_{\rm b}=0.0413$, $n_s=0.95$, $H_0 = 73$km s$^{-1}$ Mpc$^{-1}$ and $\sigma_8=0.8$, thus consistent with the 5-year analysis of the WMAP data (Komatsu et al. 2009). The simulation has a periodic box size of $75h^{-1}$ cMpc and starts with $512^3$ dark matter (DM) and gas particles each; the masses of the DM and gas particles are $\simeq 1.7\times 10^8\,h^{-1}{\rm M}_\odot$ and $\simeq 4.1\times 10^7\,h^{-1}{\rm M}_\odot$, respectively. The run assumes a metallicity-dependent radiative cooling (Sutherland \& Dopita 1993) and a uniform $z$-dependent UV background produced by quasars and galaxies as given by Haardt \& Madau (1996). The code also includes an effective model to describe self-regulated star formation in a multi-phase ISM and a prescription for galactic winds triggered by SN explosions, Springel \& Hernquist (2003). In their model, star formation occurs due to collapse of condensed clouds embedded in an ambient hot gas. Stars with mass larger then 8${\rm M_\odot}$ explode as supernovae and inject energy back into the ISM. The relative number of stars of different mass is computed for this simulation by assuming a standard Salpeter initial mass function (IMF) between 1 and 100 ${\rm M_\odot}$. Metals are produced by SNII, SNIa and intermediate and low-mass stars in the asymptotic giant branch (AGB). Metals and energy are released by stars of different masses by properly accounting for mass--dependent lifetimes as proposed by Padovani \& Matteucci (1993). The metallicity--dependent stellar yields have been taken from from Woosley \& Weaver (1995) and the yields for AGB and SNIa from van den Hoek \& Groenewegen (1997).  Galaxies are recognized as gravitationally bound groups of star particles by running a standard friends-of-friends (FOF) algorithm, decomposing each FOF group into a set of dis-jointed substructures and identifying these using the SUBFIND algorithm (Springel et al. 2001). After performing a gravitational unbinding procedure, only sub-halos with at least 20 bound particles are considered to be genuine structures, Saro et al. (2006). For each ``bona-fide'' galaxy in the simulation snapshots at $z \approx 6,7,8$, we compute the mass-weighted age ($t_*$), the total halo/stellar/gas mass ($M_h/M_*/M_g$), the SFR ($\dot M_*$), the mass weighted gas/stellar metallicity ($Z_g/Z_*$), the mass-weighted gas temperature and the half mass radius of the dark matter halo. 

We start by recapitulating the theoretical model that was used to identify LAEs using the same simulation snapshots used in this work and the interested reader is referred to Dayal et al. (2008, 2009), Dayal, Ferrara \& Saro (2010) and Dayal \& Ferrara (2011) for complete details. The SED for each galaxy in each snapshot is computed using the population synthesis code {\tt STARBURST99} (Leitherer et al. 1999), using the values of $\dot M_*$, $Z_*$ and $t_*$ of the galaxy under consideration; this includes the contribution both from stellar and nebular emission. We assume a neutral hydrogen (\HI) ionizing photon escape fraction of $f_{esc}=0.02$ (Gnedin et al. 2008) and hydrogen case-B recombination to calculate the intrinsic Ly$\alpha$ luminosity, $L_\alpha^{int}$. The intrinsic continuum luminosity, $L_c^{int}$, is calculated at rest-frame wavelengths $\lambda = 1350, 1500,1700$ \AA\, at $z \approx 6,7,8$ respectively; though these wavelengths have been chosen for consistency with observations, using slightly different values would not affect the results in any sensible way, given the flatness of the intrinsic spectrum in this short wavelength range. For each galaxy, the intrinsic Ly$\alpha$ EW is calculated as $EW^{int} = L_\alpha^{int}/L_c^{int}$.

Both the Ly$\alpha$ and continuum photons so produced are susceptible to absorption by dust grains present in the ISM; only a fraction $f_\alpha$ ($f_c$) of the Ly$\alpha$ (continuum) photons produced emerge out of the galaxy undamped by dust. Dust is produced both by SN and evolved stars in a galaxy. However, several authors (Todini \& Ferrara 2001; Dwek et al. 2007, Valiante et al. 2009) have shown that the contribution of AGB stars becomes progressively less important towards higher redshifts ($z \gsim 5.7$) since the typical evolutionary time-scale of these stars ($\geq 1$ Gyr) becomes longer than the age of the Universe above that redshift. We therefore make the hypothesis that the dust present in galaxies at $z \ge 6$ is produced solely by SNII. The total dust mass present in each galaxy is then computed assuming: (i) ${0.5\, \rm M_\odot}$ of dust is produced per SNII (Todini \& Ferrara 2001; Nozawa et al. 2007; Bianchi \& Schneider 2007), (ii) SNII destroy dust in forward shocks with an efficiency of about 40\% (McKee 1998; Seab \& Shull 1983), (iii) a homogeneous mixture of gas and dust is assimilated into further star formation, and (iv) a homogeneous mixture of gas and dust is lost in SNII powered outflows. 

To transform the total dust mass into an optical depth to UV continuum photons, $\tau$, we use the results obtained for LAEs: for each galaxy, we assume that dust is made up of carbonaceous grains and spatially distributed in a slab of radius $r_d=(0.6,1.0) r_*$ at $z \approx (6,7)$, respectively. Here, $r_*$ is the radius of the stellar distribution calculated by extrapolating the results of Bolton et al. (2008) who have derived fitting formulae to their observations of massive, early type galaxies between $z=0.06-0.36$. The continuum luminosity that reaches the observer, $L_c$, is then related to the intrinsic luminosity as $L_c = L_c^{int} f_c$. In order to achieve the best fit to the LAE Ly$\alpha$ LF, we slightly modify the relation between $f_\alpha$ and $f_c$ presented in Dayal, Ferrara \& Saro (2010): for each simulated galaxy at $z \approx (6,7)$ we use $f_\alpha = (1.5,0.6) f_c e^{-M_h/M_k}$ and $M_k$ is independent of redshift with a value $ 10^{11.7} {\rm M_\odot}$; we assume the same ratio of $f_\alpha/f_c$ at $z \approx 8$ as at $z \approx 7$. Interested readers are referred to Dayal, Ferrara \& Saro (2010) for complete details of this calculation. 

Unlike continuum photons, the Ly$\alpha$ photons that escape out of the galactic environment are further attenuated by the \HI in the intergalactic medium (IGM), and only a fraction $0 < T_\alpha <1$ reaches the observer. In turn the \HI number density at a given redshift depends on the assumed reionization history, which we have taken according to the Early Reionization Model (ERM; Gallerani et al. 2008a). Such reionization history explains a number of LAE and QSO observations (see Dayal et al. 2008; Gallerani et al. 2008ab), which are used to fix the \HI fraction $\chi_{HI} \approx (6 \times 10^{-5}, 2.4 \times 10^{-5}, 0.22)$ at $z \approx (6,7,8)$ respectively. We then compute the radius of the \HII region each galaxy is able to ionize around itself depending on its age and SFR, using $f_{esc}=0.02$ as mentioned before. However, in reality, multiple galaxies generally contribute ionizing photons to the same ionized region due to source clustering. The size of this ``collective'' \HII region is therefore larger and the photoionization rate inside it includes the direct radiation from the galaxy under consideration, those clustered around it and residing in the collective HII region, and the UV background. This procedure, described in detail in Dayal \& Ferrara (2011), was carried out for each galaxy in the simulated volume. Assuming photoionization equilibrium within the effective ionized region of each galaxy and forcing $\chi_{HI}$ to attain the assigned global value at the edge of this region, we used the Voigt profile to calculate the optical depth, and hence $T_\alpha$ for Ly$\alpha$ photons along the line of sight. The observed Ly$\alpha$ luminosity for each galaxy is then calculated as $L_\alpha = L_\alpha^{int } f_\alpha T_\alpha$.

\subsection{Identifying LAEs and LBGs}
\begin{figure*} 
   \center{\includegraphics[scale=1.0]{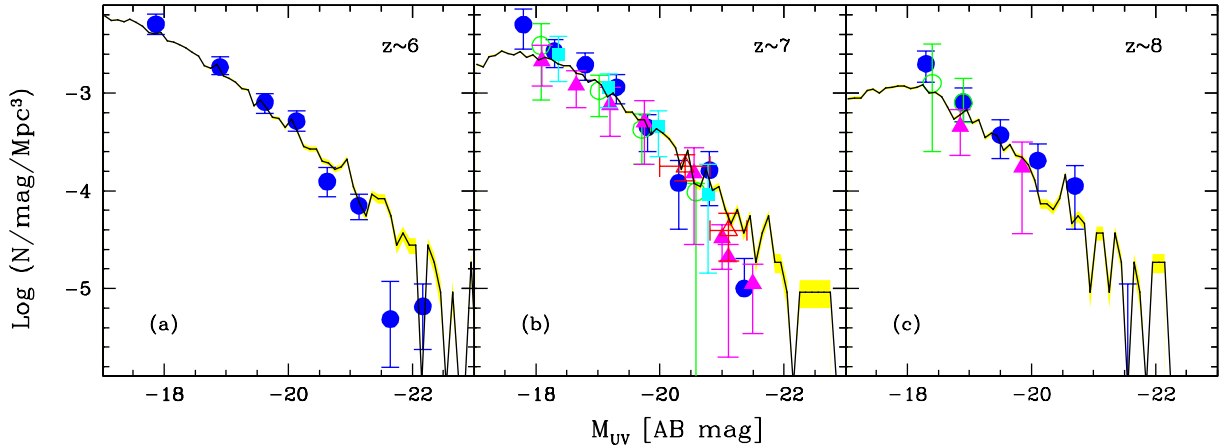}} 
  \caption{LBG UV LFs for $z \approx 6, 7, 8$ as marked in each panel. In all panels, points show the observed data, solid lines show the theoretical results and the shaded regions show the poissonian errors associated with the latter. The observed LBG UV LFs have been taken from: (a) $z\approx 6$: Bouwens et al. (2007; filled circles); (b) $z \approx 7$: Oesch et al. (2010; filled squares), Bouwens et al. (2010a; empty circles), Bouwens et al. (2010b; filled circles), Castellano et al. (2010; empty triangles) and McLure et al. (2010; filled triangles);  (c) $z \approx 8$: Bouwens et al. (2010a; empty circles), Bouwens et al. (2010b; filled circles) and McLure et al. (2010; filled triangles).}
\label{lbg_uvlf} 
\end{figure*}

When compared to the standard Sheth-Tormen mass function (Sheth \& Tormen 1999), the simulated mass function is complete for halo mass $M_h \geq 10^{10} {\rm M_\odot}$ in the entire redshift range of interest, i.e. $z \approx 6-8$; galaxies above this mass cut-off are referred to as the ``complete sample''. At each simulated redshift, $z \approx 6,7,8$, in consistency with the current observational criteria, galaxies from the ``complete sample'' with an absolute magnitude $M_{UV} \leq -17$ are identified as LBGs; galaxies with $L_\alpha \geq 10^{42.2} {\rm erg \, s^{-1}}$ {\it and} an observed Ly$\alpha$ equivalent width $EW = L_\alpha/L_c \geq 20$ \AA\, are identified as LAEs.

\section{Comparison with LBG observations}
We reiterate that galaxies identified as LAEs in this work are the same as those presented in Dayal et al. (2009) and Dayal, Ferrara \& Saro (2010). These have already been shown to successfully reproduce a number of LAE data sets including the Ly$\alpha$/continuum LFs, EW distributions, observed SEDs and the color excess, $E(B-V)$, to mention a few. We refer the interested reader to the above mentioned works for complete details. In this work, we limit ourselves to comparing the results for the theoretical LBGs to observations. We start by comparing the theoretical LBG UV LFs to the observed ones, for each of the redshifts of interest. 
 
\subsection{UV Luminosity Functions}
\label{lfs}
As mentioned before, the LBG UV LFs are by far the most solid piece of information collected for high-$z$ galaxies. Indeed, the observations by different groups are all in excellent agreement at $z \approx 7,8$, as seen from Panels (b) and (c) of Fig. \ref{lbg_uvlf}. Reproducing these LFs is therefore the first test of our model: for each redshift of interest, the theoretical LF is computed by counting the number of LBGs in each $M_{UV}$ bin and dividing it by the simulated volume, and the size of the bin. 

We find that at all the redshifts considered, both the amplitude and the slope of the theoretical LFs are in excellent agreement with the observations, over a broad magnitude range  $M_{UV} \approx -18$ to $-23$, as seen from panels (a,b,c) of Fig. \ref{lbg_uvlf}. We consider this to be an encouraging success of our model, particularly because once $L_c^{int}$ and $f_c$ are calculated using the physical properties of each galaxy, as explained in Sec. \ref{simu}, there are {\it no free parameters} left to calculate the UV LF. 

We briefly digress here to discuss the LAE UV LFs: we begin by reiterating that, as explained in Sec. \ref{simu}, the model parameters used for both LAEs and LBGs have the same values at a given redshift. The parameters used in this work have already been shown to reproduce the LAE UV LFs extremely well at $z \approx 6,7$ (see Fig. 7 of Dayal, Ferrara \& Saro 2010). Indeed this is not a surprise since observationally, both the amplitude and shape of the LAE UV LFs coincides perfectly with that observed for LBGs between $M_{UV} \approx -22$ to $-20$, at $z \approx 6,7$; this agreement has already been pointed out and discussed in detail in Shimasaku et al. (2006). Such agreement hints at the fact that LAEs and LBGs are similar galaxy populations (at least in the above UV magnitude range). 

We now return to the discussion regarding the LBG LFs where we find that the theoretical LFs shift towards lower luminosity or decreasing $M^*_{UV}$ with increasing redshift, mimicking a pure luminosity or density evolution. The value of the best-fit Schechter parameters to the theoretical LFs can be quantified by a characteristic magnitude $M^*_{UV} \approx (-20.3\pm 0.1, -20.1\pm 0.2, -19.85\pm 0.15)$ and slope $\alpha \approx (-1.7\pm 0.05, -1.7\pm 0.1, -1.6\pm 0.2)$ for $z \approx (6,7,8)$ which, within errors, are consistent with the values $M^*_{UV} \approx (-20.24\pm 0.19,-20.14\pm 0.26,-20.1\pm 0.52)$ and $\alpha \approx (-1.74 \pm 0.16, -2.01 \pm0.21, -1.91 \pm 0.32)$ derived by Bouwens et al. (2010b) for the same redshifts. 

Note, however, that the large cosmological volume simulated ($\approx 10^6 \, {\rm cMpc}^3$) naturally limits the mass resolution. As mentioned in Sec. \ref{simu}, when compared to the standard Sheth-Tormen mass function, the theoretical halo mass function is only complete for galaxies with halo mass $M_h \geq 10^{10} {\rm M_\odot}$ at any of the redshifts considered here. We are thus unable to resolve the sources populating the faint end of the LF, i.e. $M_{UV}\geq -18$, which explains the flattening of all the theoretical LFs below this magnitude limit, as shown in Fig. \ref{lbg_uvlf}. As a consequence, we are also unable to constrain the faint end slope of these LFs.  Readers interested in the faint end of the LF are referred to a previous work (Salvaterra, Ferrara \& Dayal 2011) where the authors have used higher resolution simulations with an identical implementation of physical processes, albeit with a much smaller volume ($10^3{\rm cMpc}^3$), to find an almost constant value of $\alpha \approx -2$ for the faint end slope between $z \approx 5-10$.

\subsection{UV spectral slopes}
The UV spectral slope, $\beta$, of galaxies is generally parameterized using a power law index such that the specific flux $f_\lambda \propto \lambda^\beta$ (e.g. Meurer et al. 1999); in this convention, $\beta =-2$ corresponds to a source that has a zero color in the AB magnitude system. Since $\beta$ depends on the intrinsic properties of the galaxy, it has become popular to get a hint on the values of $t_*$, $Z_*$ and $E(B-V)$ for high-$z$ LBGs. 

\begin{figure*} 
   \center{\includegraphics[scale=1.0]{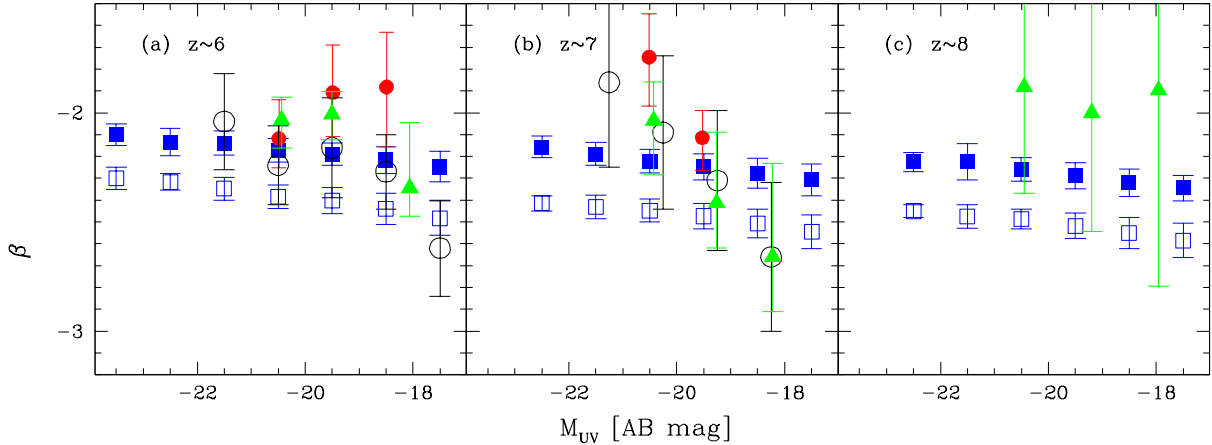}} 
  \caption{The LBG UV spectral slope, $\beta$, as a function of the UV magnitude (in bins of 1.0 dex) at $z \approx 6,7,8$, as marked in each panel. Empty (filled) squares show the theoretical $\beta$ values without (with) dust correction with the error bars showing the $1-\sigma$ error. Filled circles show the $\beta$ values inferred by Dunlop et al. (2011) for their `robust' sample, with the additional condition of at least one $8-\sigma$ near-IR detection (see Tab. 1 of Dunlop et al. 2011), empty circles show the results from Bouwens et al. (2011b), with the associated random and systematic errors and filled triangles show the $\beta$ values inferred by Finkelstein et al. (2011). }
\label{beta_spn} 
\end{figure*}

However, the values of $\beta$ inferred from broad band photometry and their evolution with redshift, have been hotly debated over the past few months as is now summarized. Using HUDF and ERS data, Bouwens et al. (2010a) find smaller $\beta$ (i.e. bluer spectra) both with decreasing magnitude and increasing redshift: LBGs at $z \approx 7$ are in general bluer than those at $z \approx 4$, and for $z \approx 7$ LBGs, $\beta$ decreases from $\approx -2$ to $\approx -3$ as $M_{UV}$ increases from $-20.5$ to $-18.5$. Dunlop et al. (2011) have used the same data set to explore the effects of including/excluding less robust LBG candidates and have considered the fact that the fields used have different limiting magnitudes, to study the effects of noise and selection bias at any given UV luminosity. These authors claim that artificially low values of $\beta$ can be found in the deepest 0.5 magnitude bin of the WFC3 selected samples, irrespective of luminosity or redshift (see also Finkelstein et al. 2010). Confining their analysis to robust LBG candidates, these authors find no trend of $\beta$ with either magnitude or luminosity: their average $\beta$ value is consistent with $\langle \beta \rangle \approx -2.05 \pm 0.1$ over $z \approx 5-7$ and between $M_{UV} \approx -22$ to $-18$. The results obtained by Dunlop et al. (2011) are supported by the recent work of Wilkins et al. (2011) who find a zero mean UV continuum color in the AB magnitude system, i.e. $\beta \approx -2$ between $M_{UV} \approx -21$ to $-18$ for WFC3-selected LBGs at $z \approx 7$. Finkelstein et al. (2011) have fit spectral synthesis models to the LBG data collected from CANDELS, the HUDF and ERS, to infer the best-fit value of $\beta$. These authors also find no trend of $\beta$ with the UV magnitude at 1500\AA\, in the rest-frame; their average $\beta$ values of $\langle \beta \rangle \approx (-2.07 ^{+0.06}_{-0.09}, -2.37^{+0.28}_{-0.06})$ at $z \approx (6,7)$ are consistent, within errors, with the results found by Dunlop et al. (2011). Recently, Bouwens et al. (2011b) have updated their results using data from HUDF09 and CANDELS data, to find $\langle \beta \rangle \approx -2.2$ at $z \approx 6,7$. These new values are now consistent with the results found by Dunlop et al. (2011, Wilkins et al. (2011) and Finkelstein et al. (2011) mentioned above.

\begin{figure*} 
   \center{\includegraphics[scale=1.0]{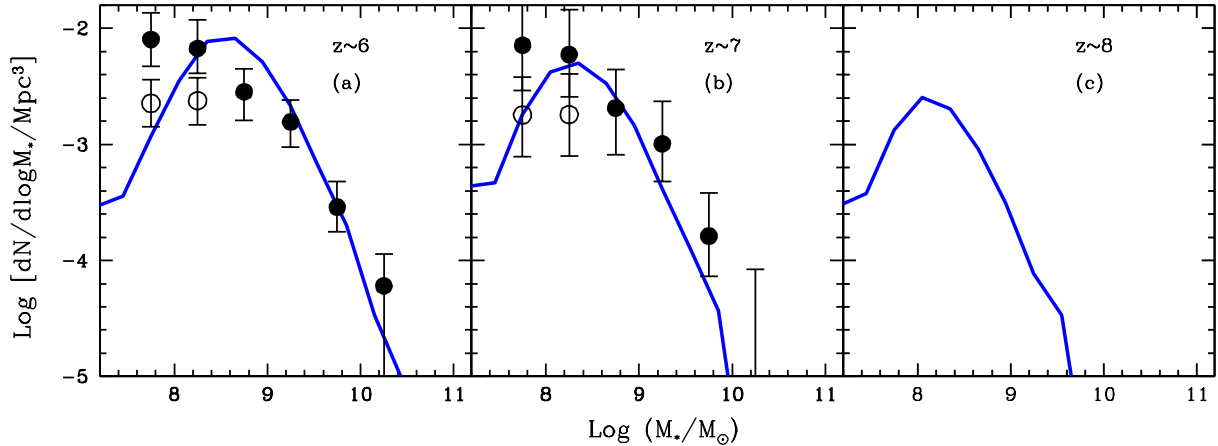}} 
  \caption{The LBG stellar mass function at $z \approx 6,7,8$ as marked in each panel. In each panel, the solid line shows the theoretical result (poissonian errors are invisible on this scale); the completeness limit of the simulation is $M_*\approx 10^{8.5} {\rm M_\odot}$. Filled (empty) points show the corrected (uncorrected) LBG stellar mass function inferred by Gonzalez et al. (2011) with error bars reflecting the uncertainty in the LF and the mass-luminosity relation used (see Gonzalez et al. 2011 for details). }
\label{mass_fn} 
\end{figure*}

We start by obtaining the intrinsic value of $\beta$ ($\beta_{int}$; i.e. without considering the effects of dust) for each of the simulated LBGs on the UV LFs at $z \approx  6,7,8$, shown in Fig. \ref{lbg_uvlf}. We use 30 values of $\beta$ evenly spaced between $-3$ and $0$ to fit a line through the intrinsic SED of each LBG (as obtained using {\tt STARBURST99}) between $1500-3000$\AA\, in the rest frame, sampled every 100 \AA; the value of $\beta$ yielding the smallest $\chi^2$-error is then chosen as the $\beta$ value for that particular galaxy. We find that $\beta_{int}$ decreases slightly for fainter galaxies at any given redshift as shown in Fig. \ref{beta_spn}. For example, $\beta_{int}$ decreases from $-2.3$ to $\approx -2.5$ as $M_{UV}$ increases from $-23.5$ to $-17.5$ at $z \approx 6$; this is due to the fact that smaller (or less luminous) LBGs are in general younger and have lower stellar metallicity, as will be discussed in more detail in Sec. \ref{phy_prop} below. Moreover, as shown in Tab. \ref{table1}, on average, the values of $t_*$ and $Z_*$ of LBGs shift to progressively lower values with increasing redshift: the average age and metallicity shift from $t_* \approx 142$ Myr ($Z_* \approx 0.14 {\rm Z_\odot}$) at $z \approx 6$ to $t_* \approx 69$ Myr ($Z_* \approx 0.05 {\rm Z_\odot}$) at $z \approx 8$. All this results in a progressively bluer SED with redshift, such that $\langle \beta_{int} \rangle \approx (-2.45,-2.5,-2.55)$ at $z \approx (6,7,8)$.

We then calculate the observed $\beta$ values, taking into account the dust enrichment of LBGs with redshift. We convolve the intrinsic spectrum of each LBG with the SN extinction curve (Bianchi \& Schneider 2007); the latter is scaled for the $f_c$ calculated for each LBG (see Sec. \ref{simu}). We use the SN curve since it has been shown to successfully interpret the observed properties of the most distant quasars (Maiolino et al. 2006) and gamma-ray bursts (Stratta et al. 2007). We again use the same fitting procedure described before to find the $\beta$ value minimizing the $\chi^2$-error. As expected, including dust attenuation makes $\beta$ redder (see Fig. \ref{beta_spn}), such that the average values are: $\langle \beta \rangle \approx (-2.2,-2.28,-2.32)$ for $z \approx (6,7,8)$. As seen from these values, the difference in $\beta$ with/without dust effects decreases with increasing redshift; this is easily explicable by the fact that galaxies are in general younger at increasing redshifts, and therefore have less time to produce dust. This argument is supported by the decreasing stellar metallicity values shown in Tab. \ref{table1}. Finally, we find that $\beta$ decreases slightly with increasing magnitude at any redshift; for example, it changes from $\approx -2.1$ to $\approx -2.25$ as $M_{UV}$ increases from $-23.5$ to $-17.5$ at $z \approx 6$. 

We now compare the above predictions with the most recent available data. We start by noting that within error bars, both the amplitude and the trend of the theoretical $\beta-M_{UV}$ nicely fits the data sets of Dunlop et al. (2011), Bouwens et al. (2011b) and Finkelstein et al. (2011) at $z \approx 6$. At $z \approx 7$, the $\beta-M_{UV}$ trend inferred observationally by these same authors appears to be steeper than the theoretical one. Such a steepening of the $\beta-M_{UV}$ relation remains hard to understand theoretically, given that the dust enrichment, ages and metallicities of both LAEs and LBGs do not evolve appreciably in $z = 6-7$ (see Sec. \ref{phy_prop} that follows). It remains to be seen whether improvements in data quality/analysis, as it has been the case at $z\approx 6$, can release such tension between theory and observations at $z=7$.

Finally, as shown in Fig. \ref{beta_spn} and discussed above, we find that $\beta$ decreases only slightly with increasing redshift/magnitude and is consistent with $\langle \beta \rangle \sim -2.2$ at $z \approx 6,7,8$. This result is consistent with all the most updated observational results available, including those of Dunlop et al. (2011), Wilkins et al. (2011), Bouwens et al. (2011b) and Finkelstein et al. (2011).

\subsection{Stellar Mass Functions}
\label{massfn}

At each redshift, the stellar mass function is calculated by counting the number of LBGs in each (logarithmic) stellar mass, $M_*$, bin and dividing it by the total simulated volume, and the size of the bin. As a first result, we see from Fig. \ref{mass_fn}  that the number density of LBGs in any given mass bin decreases with increasing redshift, as expected in standard hierarchical structure formation models, where galaxies build up their stellar mass gradually from the merger of smaller objects. Hierarchical scenarios also naturally explain the upper-mass cutoff shifting to progressively lower $M_*$ values with redshift, as the largest objects, with $M_* \approx 10^{10.2} {\rm M_\odot}$ at $z \approx 6$, have not had sufficient time to grow at $z \approx 8$, where the upper mass cut-off is $M_* \approx 10^{9.6} {\rm M_\odot}$.  

We compare these theoretical stellar mass functions to those derived by Gonzalez et al. (2011) by SED fitting of the broad band photometric colors obtained from Hubble-WFC3/IR observations combined with deep GOODS-S Spitzer/IRAC data. This is a good consistency check of our model, given that Gonzalez et al. (2011) assume a constant star formation, to find the stellar masses that best fit the observed SEDs. We find the agreement between the theoretical mass function and the data is quite satisfactory, both in terms of the slope and the magnitude of the {\it dust-corrected} estimates for stellar mass $M_*\geq 10^{8.5} {\rm M_\odot}$ at any of the redshifts considered; varying the model parameters such as modelling star formation to occur in bursts, or adding the contribution from nebular emission might affect such an agreement (see Schaerer \& deBarros 2010). Below this mass threshold, our simulations start being incomplete because of the numerical resolution issues already discussed in Sec. 2; the decrease in the mass function is probably an artifact of such resolution issues. 

\begin{figure} 
   \center{\includegraphics[scale=0.48]{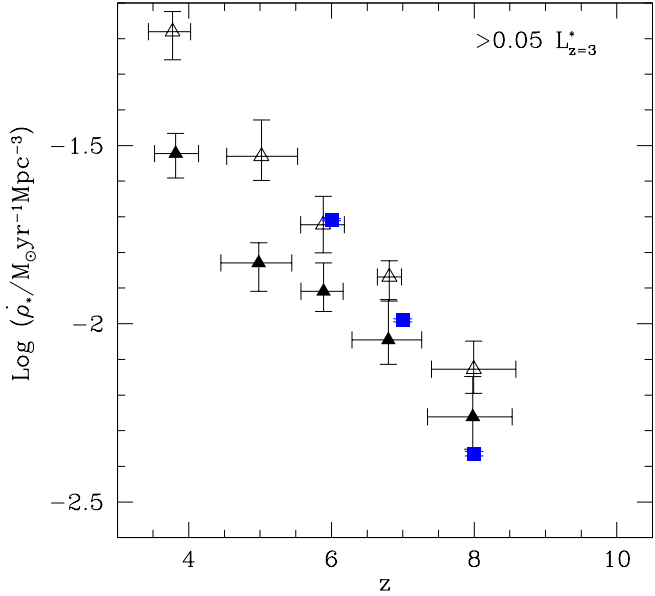}} 
  \caption{The SFR density for LBGs, obtained by integrating down the UV continuum luminosity density to 0.05$L^*_{z=3}$, plotted as a function of redshift. Filled squares show the theoretical results with error bars showing the poissonian errors; solid (empty) triangles show the dust uncorrected (corrected) values inferred by Bouwens et al. (2011b) assuming a Salpeter IMF between 0.1 and 125 ${\rm M_\odot}$ and a constant SFR for $> 100$ Myr.}
\label{sfr_den_lbg} 
\end{figure}

\begin{figure} 
   \center{\includegraphics[scale=0.48]{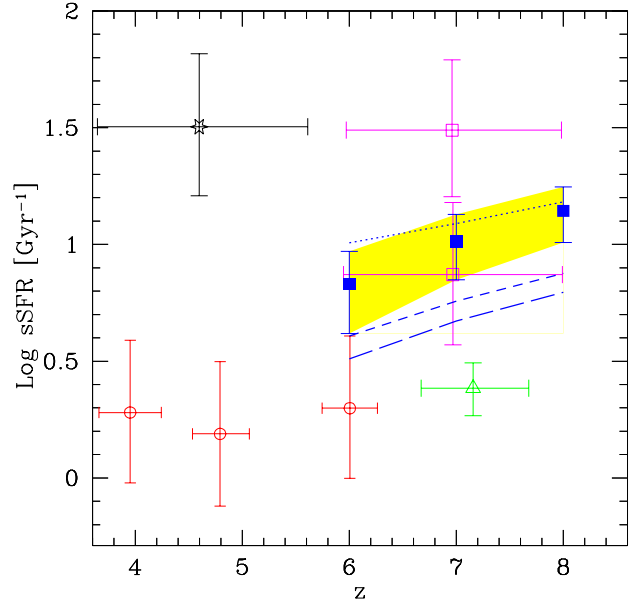}} 
  \caption{Specific SFR for LBGs plotted as a function of redshift. Filled squares show the theoretical results averaged over all stellar masses with the shaded region showing the associated poissonian errors; dotted, short and long dashed lines show the theoretical estimates for galaxies with total stellar mass $M_* < 10^{8.5} {\rm M_\odot}$, $M_* = 10^{8.5-9.5} {\rm M_\odot}$ and $M_* > 10^{9.5} {\rm M_\odot}$ respectively. Empty symbols refer to observations: the star shows the estimate of Yabe et al. (2009) for galaxies with mass between $10^{8-11} {\rm M_\odot}$ with a median mass of $4.1 \times 10^9 {\rm M_\odot}$. Circles show the estimates from Stark et al. (2009). The triangle shows the average value from Gonzalez et al. (2010) whose sample has a median mass of $5 \times 10^9 {\rm M_\odot}$. Finally, the upper and lower empty squares show the upper and lower limits inferred by Schaerer \& de Barros (2010) and refer to galaxies with masses $\approx 10^{8}$ and $10^{9.5} {\rm M_\odot}$ respectively.}
\label{ssfr} 
\end{figure}

\subsection{Star Formation Rate Densities}
Another useful quantity that can be extracted from our numerical simulations is the LBG SFR density, $\dot \rho_*$. To compare directly with available data, at each of the simulated redshifts, this value is computed by summing the intrinsic SFR of all LBGs whose observed luminosity is larger than $0.05 L^*_{z=3}$ (corresponding to $M_{UV,AB} = -21.07$ at $z=3$) and dividing this sum by the total simulated volume. We start by observing that $\dot \rho_*$ decreases with increasing redshift, tracing the cosmic SFR density (e.g. see Tab. 2 of Hopkins 2004). Further, as seen from Fig. \ref{sfr_den_lbg}, the theoretical SFR density values are in good agreement with the dust-corrected (uncorrected) estimates of Bouwens et al. (2011b) for $z \approx 6$ ($z \approx 7,8$). 

This agreement,  however, is probably the result of an IMF-dust attenuation degeneracy as we explain in the following. Bouwens et al. (2011b) have used a Salpeter IMF between 0.1-125 ${\rm M_\odot}$ and $f_c \approx (0.66,0.77,0.77)$ at $z \approx (6,7,8)$ respectively, as seen from a comparison of their dust corrected/uncorrected points at these redshifts (see Tab. 7, Bouwens et al. 2011b). Our simulations instead, use a Salpeter IMF between $1-100 {\rm M_\odot}$; the IMF lower limit of $1 {\rm M_\odot}$ results in an intrinsic luminosity which is about 2.5 times larger compared to that from the IMF used by Bouwens et al. (2010b). As expected, to fit the UV LF amplitude, our $f_c \approx (0.2,0.3,0.36)$ values at $z \approx (6,7,8)$ (corresponding to a mean $E(B-V) \approx (0.16,0.12,0.1)$, see Tab. \ref{table1}) are then about 2.5 times smaller than those of Bouwens et al. (2010b). Thus, although theoretical dust corrections also decrease with increasing redshift, they are {\it not negligible} even in the most distant LBGs (see also Schaerer \& de Barros 2010 and Laporte et al. 2011).

\begin{figure*} 
  \center{\includegraphics[scale=0.9]{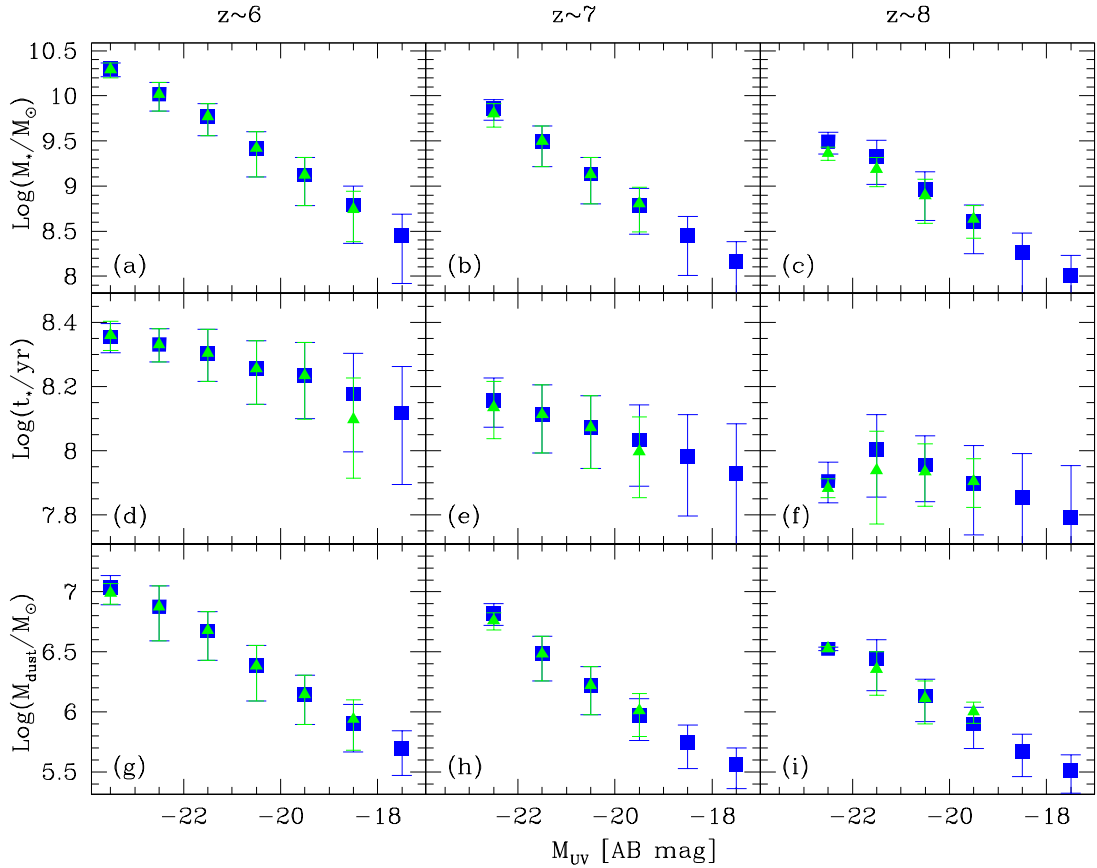}} 
  \caption{As a function of $M_{UV}$ (in bins of 1 dex), we show the physical properties of LBGs (squares) and LAEs (triangles). The physical quantities represented by the rows are (a) {\it top row:} the total stellar mass, (b) {\it middle row:} the mass weighted stellar age and (c) {\it bottom row:} the total dust content. Columns from left to right refer to $z \approx 6,7,8$ as marked and the error bars show $1-\sigma$ errors.  }
  \label{phy} 
\end{figure*}

We now turn to our predictions for the specific SFR, where for each galaxy, the sSFR is calculated as ${\rm sSFR}= \dot M_*/M_*$. At each of the simulated redshifts, the value of sSFR in a given $M_*$ bin is the average over all the LBGs in that mass bin. First, we find that the sSFR decreases with increasing $M_*$; galaxies with $M_* = 10^{8.5-9.5} {\rm M_\odot}$ and $M_* > 10^{9.5} {\rm M_\odot}$ show less than half of the sSFR of galaxies with $M_* < 10^{8.5} {\rm M_\odot}$, at any of the redshifts considered here. This apparently puzzling result might be the smoking gun of downsizing: whereas more massive galaxies formed large amounts of stars in a short time interval which quenched later star formation, dwarf galaxies built up their stellar budget more gradually. Thus, by the time the SFR is roughly similar for the two populations, more massive galaxies already contain a large amount of stellar mass, and are therefore characterized by a lower sSFR. This trend appears to hold at all redshifts, as seen from Fig. \ref{ssfr}. We note that our estimates of the sSFR of galaxies with $M_* = 10^{8.5-9.5} {\rm M_\odot}$ are consistent with the those obtained by Schaerer \& de Barros (2010) for galaxies with $M_* = 10^{9.5} {\rm M_\odot}$ (Fig. \ref{ssfr}). Further, averaged over all LBGs at a given redshift, the sSFR rises with redshift, from $\approx 6.7$ to $\approx 13.9\, {\rm Gyr^{-1}}$ from $z \approx 6$ to $z \approx 8$, as shown in Fig. \ref{ssfr} and quantified in Tab. \ref{table1}. Such a trend is easily explained combining the fact that galaxies shift to progressively lower halo (and stellar) masses with redshift (see Fig. \ref{mass_fn}) and as mentioned above, the sSFR rises with decreasing stellar mass.   

\section{LAE-LBG connection}
\label{phy_prop}
Having identified the simulated galaxies that would be observationally selected as LBGs, and shown that they reproduce a number of observed data sets, we can now compare their properties with those of the LAEs previously found in the same simulations. 

We start by considering the stellar mass, $M_*$, which is possibly the best constrained physical quantity available for high-$z$ galaxies when near-IR data is available. We find that for galaxies brighter than  $M_{UV} \leq (-18, -19, -19)$ at $z \approx (6,7,8)$, LAEs and LBGs show very similar values of $M_*$ in a given $M_{UV}$ bin.  The stellar mass and UV magnitude are tightly coupled at each redshift; galaxies with the largest $M_*$ (and hence $\dot M_*$; see Fig. 7 of Dayal et al. 2009) are the most UV luminous, implying that although dust obscuration can modify the slope of the SFR-UV luminosity relation, it cannot alter its monotonic trend. A persistent evolutionary feature is that the LBG population extends to lower stellar masses/fainter luminosities: this is because the smallest galaxies do not produce enough Ly$\alpha$ to be visible as LAEs, according to the adopted selection criterion ($L_\alpha \geq 10^{42.2} {\rm erg \, s^{-1}}$ and $EW \geq 20$ \AA). Finally, the $M_*$ ranges shift to progressively lower values with redshift (see also Sec. \ref{massfn}) for both LAES and LBGs, as expected in a hierarchical structure formation scenario; at $z \approx 6$ the stellar masses for LBGs (LAEs) range between $10^{8.5-10.3} (10^{8.7-10.3}) {\rm M_\odot}$, while they shift down to $M_* \approx 10^{8-9.5} (10^{8.7-9.4}) {\rm M_\odot}$ at $z \approx 8$ (Fig. \ref{phy}). The LAE stellar mass range evolves faster than that for LBGs, due to the combined effects of dust and reionization, as explained in Sec. \ref{reio} below. 

Similar considerations hold for the stellar ages, $t_*$, of the two populations, as they are found to be very similar in any given absolute magnitude bin for $M_{UV} \leq (-18, -19, -19)$ at $z \approx (6,7,8)$, further pointing to the tight relation between these two types of galaxies. Both LBGs and LAEs are composed of intermediate age systems, with $100 < t_*/{\rm Myr} < 250$ ($60 < t_*/{\rm Myr} < 250$) at $z \approx 6$ (8), as seen from panels (d)-(f) of Fig. \ref{phy}. As expected, the average $t_*$ decreases with redshift; in agreement with the downsizing scenario advocated above, larger systems are more dominated by older populations than smaller systems. Finally, the galaxies brightest in the UV are on average the oldest, by virtue of their largest $M_*$ values.

\begin{figure*} 
   \center{\includegraphics[scale=1.0]{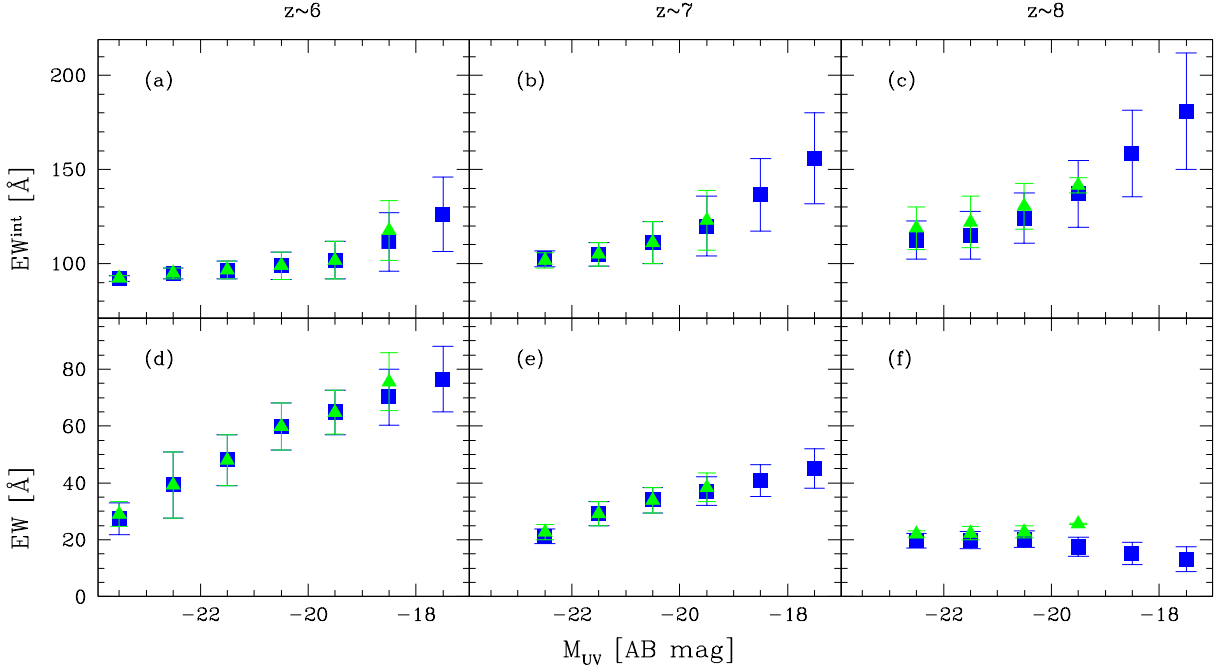}} 
  \caption{As a function of $M_{UV}$ (in bins of 1 dex), we show the intrinsic (top three panels) and the observed Ly$\alpha$ EW (bottom three panels). The values for LAEs (triangles) and LBGs (squares) are shown, with error bars referring to $1-\sigma$ errors. Columns refer to results for $z \approx 6,7,8$ as marked.}
\label{ewint} 
\end{figure*}

Finally, as shown in panels (g)-(i) of Fig. \ref{phy}, both LAEs and LBGs have nearly identical dust masses (again, apart from the faintest magnitude bins), as a result of their overlapping stellar masses and ages. As  galaxies are on average younger and less massive at higher redshifts, more distant objects tend to be less dusty such that the dust mass, $M_{dust} = 10^{5.7-7} (10^{5.9-7}) {\rm M_\odot}$ at $z \approx 6$ and $M_{dust} = 10^{5.5-6.5} (10^{6.1-6.5}) {\rm M_\odot}$ at $z \approx 8$ for LBGs (LAEs), respectively. Finally, as expected, the dust content of galaxies increases with $M_*$ (decreasing $M_{UV}$); the brightest galaxies have formed the largest amount of stars over a longer period of time compared to smaller systems. 

From the above discussion, we conclude that high-redshift sources identified as LAEs and LBGs essentially sample the same underlying galaxy population; however, LAEs represent a {\it subset} of a larger LBG sample in the sense that only the faintest LBGs do not show an observable Ly$\alpha$ line. We clarify that LAEs are neither younger/less dusty/smaller nor older/more dusty/larger as compared to LBGs, but rather a wide range of LBGs show an observable Ly$\alpha$ line. Indirectly, our findings are supported also by the recent results of Malhotra et al. (2011) who find similar physical sizes for LAEs and LBGs at $z \geq 5$. Intriguingly, studying 92 $\langle z \rangle = 2.65$ UV continuum selected galaxies to very low surface brightness limits, Steidel et al. (2011) find that extended, diffuse Ly$\alpha$ emission increases the total Ly$\alpha$ flux by a factor of about 5 on average; accounting for such diffuse emission, essentially all their LBGs would qualify as LAEs.

Finally, we caution the reader that the overlap between the LAE and LBG populations depends sensitively on redshift dependent observational selection criterion such as the UV magnitude, Ly$\alpha$ flux and EW limits, used to select these two populations; the results found in this work which suggest LAEs to be a subset of LBGs might not necessarily hold true for data sets at other redshifts selected using different selection criterion. This point has already been made in a number of previous works: Ouchi et al. (2008) have shown that the fraction of LBGs showing a Ly$\alpha$ line depends on the Ly$\alpha$ EW cut imposed at any given redshift, Verhamme et al. (2008) have shown that while LAEs brighter than a certain limiting magnitude (which is a function of redshift) are the same as LBGs, at fainter magnitudes, LAEs are less massive compared to LBGs. Also, Dijkstra \& Wyithe (2011) have pointed out that simulating LAEs using simple cuts in EW and UV magnitude lead to uncertainties in the predicted LAE number densities. 
 
\section{Imprints of reionization}
\label{reio}
%
\begin{figure*} 
  \center{\includegraphics[scale=1.0]{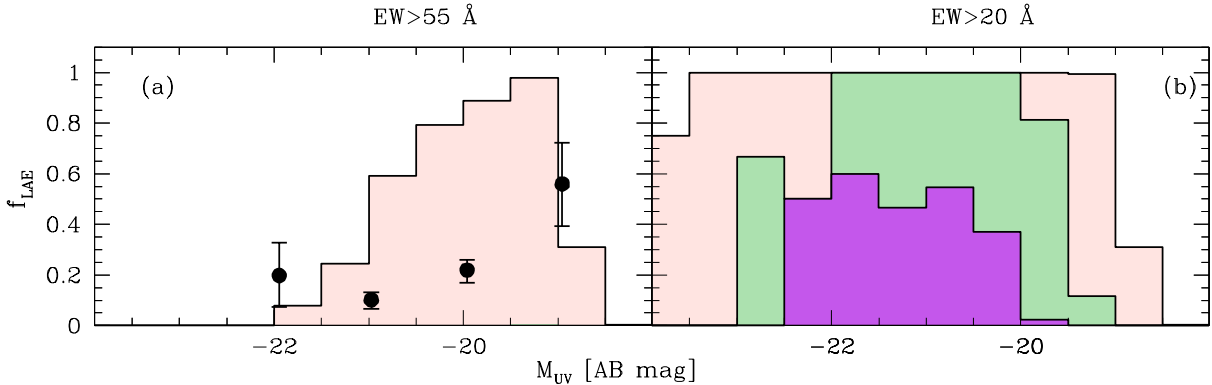}} 
  \caption{As a function of $M_{UV}$ (in bins of 1 dex), we show the fraction of LBGs showing a Ly$\alpha$ line. The left and right panels refer to the fraction of LBGs showing a Ly$\alpha$ line with $EW > 55$\AA\, and $EW>20$\AA\,, respectively. The points in the left panel show the fractions obtained by Stark et al. (2010) in the range $3 < z < 6$ and the histogram shows the model results for $z \approx 6$; the results for $z \approx 7,8$ are not visible on this plot. In the right panel, histograms from the widest to the narrowest show the model results for $z\approx 6,7,8$, respectively. See text for details.}
\label{rat} 
\end{figure*}

As explained in Sec. \ref{simu}, while both Ly$\alpha$ and continuum photons are attenuated by ISM dust, the IGM ionization state only affects the observed Ly$\alpha$ luminosity, leaving the observed continuum luminosity unchanged; a decrease in the observed EW with increasing redshift could then possibly indicate an increasingly neutral IGM, a test proposed by various authors (e.g. Stark et al. 2010). We now discuss both the intrinsic and observed Ly$\alpha$ EWs for the LAEs and LBGs identified at each of the simulated redshifts. 

As discussed in Sec. \ref{phy_prop}, for a given $M_{UV}$ value, both $t_*$ and $Z_*$ shift to progressively lower values with increasing redshift, for both LAEs and LBGs. Since the stellar spectra of younger, more metal-poor populations are harder (i.e. stars produces more \HI ionizing photons) the value of $EW^{int}$ increases with redshift for any UV magnitude bin; for the same reasons, fainter galaxies exhibit larger $EW^{int}$ values at all redshifts. Such trends hold for both LAEs and LBGs by virtue of their nearly identical physical properties, as shown in panels (a)-(c) of Fig. \ref{ewint}. Quantitatively, the range of $EW^{int} \approx 90-130$\AA\, (brightest to faintest galaxies) at $z \approx 6$ and increases to $\approx 110-180$ \AA\, at $z \approx 8$. 

Before we discuss the observed EWs, we recall that the following relation holds: $EW = EW^{int}(f_\alpha/f_c)T_\alpha$, where $T_\alpha$ is the transmissivity of Ly$\alpha$ photons through the IGM. At $z \approx 6-7$, according to the Early Reionization Model that best fits the observations, the \HI fraction is $\chi_{HI} \approx 10^{-5}$; this largely ionized IGM results in transmissivity values constrained in the narrow range $T_\alpha\approx 0.4-0.6$, from the smallest to the largest LAEs (see also Fig. 3, Dayal et al. 2009). At $z \sim 8$, instead, the more neutral IGM ($\chi_{HI} \approx 0.22$) substantially and preferentially damps Ly$\alpha$ photons from the smallest galaxies which therefore become undetected in narrow band observations; the transmissivity decreases globally to $T_\alpha \approx 0.05-0.4$. 

Further, we remind the reader that have used $f_\alpha/f_c = (1.5,0.6,0.6) e^{-M_h/M_k}$ at $z \approx (6,7,8)$, where $M_h$ is the galaxy halo mass and $M_k = 10^{11.7}{\rm M_\odot}$, as required to best reproduce the amplitudes of the cumulative LAE Ly$\alpha$ LFs. When taken together, these facts result in the value of $EW$ rising smoothly from $\approx 30$\,\AA\, at $M_{UV} \approx -23.5$ to $EW \approx 80$\,\AA, at $M_{UV} \approx -17.5$, at $z \approx 6$. However, the value of $EW$ is much flatter at $z \approx 7$ as a result of the lower escape fraction of Ly$\alpha$ photons from the ISM, compared to that for the continuum photons ($f_\alpha/f_c = 0.6$) required to match the LAE LFs.  Finally, at $z \approx 8$, the trend reverses, with the smallest galaxies showing the smallest $EW$, as a result of their strongly reduced Ly$\alpha$ transmission through a partially neutral IGM.

Given the above discussion, is then clear that reionization is the key factor (but by no means the only one) regulating the relative fraction of observed
LBGs and LAEs. With our results at hand, it is straightforward to predict the fraction of LBGs that show a visible Ly$\alpha$ line (i.e. $L_\alpha \geq 10^{42.2} {\rm erg \, s^{-1}}$), at any given redshift. An observational consensus on this quantity is far from being reached, as it will be clear from the following.  For example, using Keck spectroscopy of about 600 LBGs in the range $3 < z < 6$, Stark et al. (2010) find that the least luminous LBGs most often show Ly$\alpha$ emission with $EW > 55$\AA. To compare to such observations, we select all LBGs that would be identified as LAEs with $EW > 55$\AA, at each of the redshifts considered. We find that at $z \approx 6$, indeed it is the least luminous LBGs most often show Ly$\alpha$ emission as shown in panel (a) of Fig. \ref{rat}; qualitatively our results broadly agree with those of Stark et al. (2010). However, at higher redshifts we find only extremely few ($z=7$) or no ($z=8$) LAEs with $EW >55$\AA\, due to the low escape fraction of Ly$\alpha$ and increasing IGM opacity, as explained above. These combined effects lead to a negligible fraction of LAEs being found amongst $z > 6$ LBGs in our model.

However, the most often adopted EW cut corresponds to $EW > 20$\AA. Using this value, we find that within the range $M_{UV} \approx -23$ to $-20$, all LBGs would be classified as LAEs at $z \approx 6$. This is perhaps not surprising, given that the amplitudes of both the LAE and LBG UV LFs match perfectly within this range, as has already been pointed out by Shimasaku et al. (2006) and discussed in detail by Ouchi et al. (2008). The magnitude range of LBGs showing a Ly$\alpha$ line decreases with redshift: the smallest galaxies do not even produce enough Ly$\alpha$, as mentioned before; the fraction of the largest LBGs showing Ly$\alpha$ emission progressively decreases with redshift due to the dependence of $f_\alpha$ on $M_h$ explained above. Further, the decrease in the fraction of LBGs showing Ly$\alpha$ emission between $z \approx 6-8$ shown above, has recently also been found in observational data by both Pentericci et al. (2011) and Schenker et al. (2011); these authors find that the fraction of LBGs showing Ly$\alpha$ emission drops between $z \approx 6-7$, which they interpret as a sign of the IGM becoming progressively neutral (see also Fontana et al. 2010 and Vanzella et al. 2011). We reiterate that although we see such a trend, between $z \approx 6-7$ it is driven predominantly by the dust distribution (which governs the ratio $f_\alpha/f_c$); between $z \approx 7-8$ on the other hand, this trend arises due to a combination of the dust distribution, and the IGM being more neutral.  

As a final note of caution, we restate that results regarding the Ly$\alpha$ EWs depend sensitively on a number of assumptions and modeling of poorly understood physical processes, namely: (a) we have used a luminosity-independent value of $f_\alpha/f_c$, for all galaxies at a given redshift. This is empirically fixed by normalizing the theoretical LAE LFs to the observed ones. While a number of works (Neufeld 1991, Hansen \& Oh 2006) have emphasized that such ratio depends on the dust distribution (i.e. homogeneous/clumped) in galaxies, at the moment, essentially no experimental hint is available at high-$z$ on this aspect;  (b) the effects of peculiar velocities in calculating $T_\alpha$ have not been included; inflows (outflows) blueshift (redshift) the Ly$\alpha$ line, thereby decreasing (increasing) $T_\alpha$. A further complication arises due to the fact that both these above mentioned effects depend on the line of sight between the LAE and the observer. Finally, we note that we have used the stellar mass weighted age and assumed the SFR to be constant over such time period to compute the intrinsic spectrum of each galaxy. However, using the complete star formation history of each galaxy might affect the results presented here. Each of these points is discussed in more detail in the next Section. Alternatively, the tension between the model and the observations could also arise due to poorly understood biases/selection effects present in the data.

\section{Summary and discussion}
\label{conc}

We have extended our previous studies aimed at modeling Lyman Alpha Emitters (LAEs) to the second major population of  
high-$z$ sources, Lyman Break Galaxies (LBGs), with the final goal of investigating the physical relationship 
between them. In a set of large ($\approx 10^6 {\rm cMpc}^3$) cosmological SPH simulations, including a detailed 
treatment of star formation, feedback, metal enrichment and supernova dust production, we identify LBGs 
as galaxies with an absolute magnitude $M_{UV} \leq -17$, in consistency with current observational criteria; the same simulations 
have already been shown to match a number of observed data sets for LAEs (Dayal et al. 2010). For the LBGs identified at each of the redshifts of interest, i.e. $z \approx 6,7,8$, we compute the evolution of their (a) UV LF, 
(b) UV spectral slope, $\beta$, (c) stellar Mass Function, (d) SFR density, (e) sSFR, and compare them 
with available data. The main results from this study are summarized as follows.

\begin{itemize}
\item{The shape and amplitude of the observed LBG UV LFs at $z \approx 6,7,8$ are reproduced extremely well without needing to invoke any free parameters in addition to those used to model the LAE Ly$\alpha$/UV LFs; the best-fit Schechter function values obtained are $M^*_{UV} \approx (-20.3\pm 0.1, -20.1\pm 0.2, -19.85\pm 0.15)$ and $\alpha \approx (-1.7\pm 0.05, -1.7\pm 0.1, -1.6\pm 0.2)$ at $z \approx 6,7,8$, respectively.}

\item{The mean intrinsic LBG UV spectral slope shows little variation, either with magnitude (in $M_{UV} \approx -22.5$ to $-17.5$) or redshift, such that $\langle \beta^{int} \rangle = (-2.45, -2.5, -2.55)$ at $z \approx (6, 7, 8)$. When convolved with the SN extinction curve, the observed spectral slope increases such that $\langle \beta \rangle = (-2.2, -2.28, -2.32)$ in the same redshift range, again varying only slightly between the brightest and faintest galaxies. This is in line with the recent findings of Dunlop et al. (2011), Wilkins et al. (2011), Bouwens et al. (2011b) and Finkelstein et al. (2011).}

\item{The LBG stellar mass function matches both the slope and amplitude of the values inferred from observations by Gonzalez et al. (2010) for $M_* \geq 10^{8.5}{\rm M_\odot}$ at all of the redshifts considered here; below this mass cut-off, numerical resolution limits the accuracy of our prediction.}

\item{The LBG sSFR rises with redshift such that sSFR$ \approx 6.75$ ($13.9$) Gyr$^{-1}$ at $z \approx 6$ ($8$), with the smallest galaxies showing the largest sSFR values. We have suggested that such behavior could be a manifestation of downsizing. Finally, our results agree well with the dust-corrected SFR density values inferred by Bouwens et al. (2011b) for $z = 6-8$. }

\item{As for the LAE-LBG connection, at $z \approx 6, 7$, the LAE and LBG UV LFs have the same amplitude and slope between $-22 < M_{UV}  < -20$; within this magnitude range all LBGs would be identified as LAEs and vice-versa, according to the selection criteria usually adopted (see Sec. 2.1). However, the LBG magnitude range is more extended towards fainter luminosity at all redshifts considered. It follows that LAEs represent a luminous LBG  {\it subset}, whose relative extent depends only on the adopted selection criteria.}

\item{In addition to the previous point, for $M_{UV} \leq -18, -19$ at $z \approx 6, 7/8$ respectively, LBGs and LAEs share very similar values of $M_*, t_*$ and $M_{dust}$, in any given magnitude bin. This further strengthens the case for the argument that LAEs and LBGs are essentially the same class of galaxies; the differences inferred between these two populations solely lie in the different techniques employed for their identification. }

\item{In broad agreement with the results of Stark et al. (2010), the faintest LBGs most often show Ly$\alpha$ emission {\it if and only if} the line EW selection criterion is made more stringent, i.e. $EW>55$\AA. If instead the threshold $EW>20$\AA, is used as per the canonical criterion used for identifying LAEs, we find that all LBGs within the magnitude range $M_{UV} \approx -23$ to $-20$ would be identified as LAEs at $z \approx 6$. Further, the UV range (fraction of LBGs showing Ly$\alpha$ emission) narrows (decreases) with increasing redshift. }

\end{itemize}

In spite of the remarkable predictive power of the theoretical model presented, considerable room for improvements remains. In fact a number of delicate, albeit reasonable, assumptions have been made as explained throughout the text, which are now summarized. First, the relative escape fraction of Ly$\alpha$ to continuum photons, $f_\alpha/f_c$, used is essentially constant at a given redshift with a tiny halo mass dependence, that has been fixed by matching the amplitude of theoretical LAE Ly$\alpha$ LFs to observations. Such ratio sensitively depends on the dust distribution (homogeneous/clumpy) in the ISM of a galaxy. While all existing extinction curves (SN, Calzetti, Milky Way) assume dust to be homogeneously distributed to find $f_\alpha/f_c <1$, the ratio can be boosted to $f_\alpha /f_c >1$ if dust is clumped in the ISM, as shown by Neufeld (1991) and Hansen \& Oh (2006). Modeling of this process is beyond the capabilities of the present study since the large cosmological volumes required to achieve sufficient galaxy population statistics do not allow us to treat fine ISM details. 

Secondly, the observed Ly$\alpha$ luminosity also depends on the peculiar velocities which have not been included in our calculation; inflows/outflows into/from the galaxy can blueshift/redshift the Ly$\alpha$ line, thereby decreasing/increasing $T_\alpha$ (Verhamme et al. 2006; Dayal, Maselli \& Ferrara 2011). However, the extent to which peculiar velocities influence $T_\alpha$ is debatable since these calculations have mostly been performed under idealized situations. For example, Verhamme et al. (2006) have used spherically symmetric outflows of \HI to show an enhancement in $T_\alpha$; however, many studies, e.g. Fangano et al. (2007) and references therein, have shown that Kelvin-Helmholtz instabilities would result in breaking-up such symmetric outflows. 

Thirdly, both observationally and theoretically, the dependence of $f_{esc}$ on the halo mass and redshift have been hugely debated with the values obtained by different groups ranging between $f_{esc} = 0.01 - 0.8$ (e.g. Ricotti \& Shull 2003; Gnedin et al. 2008; Wise \& Cen 2009; Razumov \& Larsen 2010; Vanzella et al. 2010; Nestor et al. 2011). We have used a constant value of $f_{esc}=0.02$ for all galaxies in each of the simulation snapshots, following the results of Gnedin et al. (2008) who find $f_{esc}=0.01-0.03$ for galaxies with mass $\geq 10^{11} {\rm M_\odot}$ at $z \approx 3-5$. We remind the reader that $f_{esc}$ is a crucial parameter that enters into our calculations, and an increase (decrease) in its value would lead to a corresponding decrease (increase) in the nebular recombination line flux, thereby decreasing (increasing) both $L_\alpha^{int}$ and $L_c^{int}$ while increasing (decreasing) the size of the Str\"omgren region built by any galaxy. 

Fourthly, due to simulation constraints, we have used the stellar mass weighted age and assumed the SFR to be constant over such time period to compute the intrinsic spectrum of each galaxy. However, using the complete star formation history of each galaxy, i.e. using the age, metallicity and mass of each individual star particle in each galaxy to build the composite spectrum, might affect the results presented here, which is the focus of an ongoing study.

All of these above-mentioned physical effects affect the EW calculations presented in this paper; the EW comparison with observations (especially with those of Stark et al. 2010 carried out in this work) must therefore be seen only as very tentative and preliminary. It is hoped that the increasing complexity and power of numerical simulations, and upcoming instruments such as the JWST, ALMA and ongoing surveys with the HST/Keck will help disentangling such issues. 
 
\section{Acknowledgements}
The authors thank S. Borgani, L. Tornatore and A. Saro for providing the simulations used in this work. We thank R. Bouwens, M. Castellano, J.S. Dunlop, J. Forero-Romero, B.K. Gibson, L. Pentericci and D. Schaerer for stimulating discussions. We acknowledge the `First galaxies at Ringberg' workshop for numerous lively discussions about this work. PD thanks SISSA, Trieste, for their generous allocation of cluster computing time, through which a part of these calculations were carried out.


\end{document}